\documentclass[prd,twocolumn,superscriptaddress,nofootinbib,floatfix]{revtex4-2}
\usepackage{amsmath,amssymb,graphicx,booktabs,hyperref,xcolor}
\hypersetup{colorlinks=true,linkcolor=blue!60!black,citecolor=blue!60!black,urlcolor=blue!60!black}
\newcommand{\dd}{\mathrm{d}}
\newcommand{\Edot}{\dot E}
\newcommand{\sA}{\mathsf A}
\newcommand{\sC}{\mathsf C}
\newcommand{\sF}{\mathsf F}
\newcommand{\Ah}{\hat{\mathcal A}}
\newcommand{\Bh}{\hat{\mathcal B}}

\begin{document}

\title{Gravitational radiation from a photon on the light ring:\\ multipole fluxes and the logarithmic divergence}

\author{Enrico Barausse}
\affiliation{SISSA, Via Bonomea 265, 34136 Trieste, Italy}
\affiliation{INFN Sezione di Trieste, Via Valerio 2, 34127 Trieste, Italy}
\affiliation{IFPU -- Institute for Fundamental Physics of the Universe, Via Beirut 2, 34014 Trieste, Italy}

\date{\today}

\begin{abstract}
A massless particle on the circular null geodesic of a Schwarzschild or Kerr black hole radiates gravitational waves at all multipoles $\ell$ with a power that falls off only as $1/\ell$, so that the total instantaneous power is logarithmically divergent. We derive this result analytically from the Zerilli--Regge--Wheeler and Teukolsky equations with a null stress-energy tensor as source, and find that both the power radiated to infinity and the power absorbed by the horizon approach $\Edot_\ell\to g(a)\,\kappa E^2/(M^2\ell)$, where $E$ is the energy of the particle, $M$ the mass of the black hole and $a$ its spin, with $\kappa=0.063653$ and a spin factor $g(a)$, for which we obtain, for any spin, the closed form $g=27M^2(r_0-M)/[r_0^2(r_0+3M)]$ in terms of the radius $r_0$ of the Kerr light ring. The $1/\ell$ behavior has two causes: the near-resonance of the photon with the quasinormal modes, which prevents any exponential suppression at large $\ell$, and a cancellation in the source that is specific to massless particles, which reduces the power per multipole by a factor $\ell^2$ with respect to a generic source. At leading order half of the radiation falls into the black hole, and we compute the first correction to this equal split in the non-spinning case. The same analysis describes massive particles on orbits close to the light ring, where the mass turns the $1/\ell$ law into an exponential cut-off. We recover the dominant, even-parity term of the geodesic synchrotron radiation formulae for massive particles of Breuer, Ruffini, Tiomno and Vishveshwara (1973), but we find that their odd-parity term is too small by a factor of four. All results are checked against numerical solutions, up to $\ell=12800$ in Schwarzschild and $\ell=800$ in Kerr.
\end{abstract}

\maketitle

\section{Introduction}

The point-particle approximation is widely employed in black-hole (BH) perturbation theory. A small body of mass $\mu$ moving in the field of a BH of mass $M\gg\mu$ is treated as a point source for the linearized Einstein equations, and the gravitational-wave (GW) fluxes that it emits are computed perturbatively in $\mu/M$, by now through second order \cite{Pound:2019lzj,Warburton:2021kwk,Wardell:2021fyy}. This is how the signals of extreme mass-ratio inspirals, one of the main targets of the Laser Interferometer Space Antenna (LISA), are modeled, and the approach turns out to be remarkably accurate even when extrapolated to comparable masses (see e.g.\ Refs.~\cite{LeTiec:2011bk,LeTiec:2014oez,vandeMeent:2020xgc,Rifat:2019ltp,Wardell:2021fyy} and references therein). The fluxes that it yields are normally finite. A notable exception is the instantaneous flux from a body on a circular orbit near the light ring, the unstable circular null geodesic that sits at $r=3M$ in Schwarzschild: as the orbit approaches the light ring, the flux diverges while the energy of the body stays finite. This divergence was established numerically in Ref.~\cite{Barausse:2021ryv}, but an analytic proof was lacking.

The GW emission from massive particles on unstable circular orbits close to the light ring has a long history. It was studied intensively in the early '70s, under the name of geodesic synchrotron radiation \cite{Misner:1972jf,Misner:1972kg,Davis:1972dm,Chitre:1972fv,Breuer:1973kt,Breuer:1973pv,Breuer:1975LNP,Breuer:1973bchm,Hughes:1973ap,Chrzanowski:1974zb,Ruffini:1973}, after Misner proposed that the strong beaming of this radiation could reconcile the GW bursts reported by Weber \cite{Weber:1969,Weber:1970} with a source at the Galactic center \cite{Misner:1972jf}. That interpretation did not survive, but the calculations left behind a remarkable result about the shape of the spectrum. A massive particle on a circular orbit at $r_0=3M(1+\delta)$ in Schwarzschild has energy per unit rest mass $\gamma\equiv E/\mu\simeq(9\delta)^{-1/2}$: the closer the orbit is to the light ring, the more relativistic the particle. Its power spectrum scales with the frequency as $\omega^{-1}$ up to an exponential cut-off at $\omega_{\rm crit}\propto\gamma^2=E^2/\mu^2$, i.e.\ it carries equal power in each decade of frequency up to a frequency that, at fixed energy $E$, is set by the rest mass $\mu$ \cite{Breuer:1973kt,Chitre:1972fv}.

Taken at face value, this spectrum implies a divergence in the massless limit $\mu\to0$ at fixed $E$, i.e.\ $\delta\to0$: the cut-off goes to infinity and the total power at fixed energy of the particle grows without bound. That limit, however, was never taken explicitly.

In Ref.~\cite{Barausse:2021ryv} we revisited the question of whether the flux from a particle exactly on the light ring is finite or not, with numerical solutions of the Teukolsky equation extending to $\ell\sim400$. The answer was that the flux per multipole approaches $\kappa E^2/(M^2\ell)$ with $\kappa\simeq0.064$, both at infinity and at the horizon, so that the sum over multipoles diverges logarithmically. The slow convergence of mode sums near the light ring had also been noticed in the self-force literature \cite{Akcay:2012ea,Gundlach:2012aj}, and in Ref.~\cite{Barausse:2021ryv} we suggested that relativistic beaming was responsible for it, while noting that we had no formal proof. The purpose of this paper is to prove the divergence, to obtain the coefficient $\kappa$ in closed form, and to identify its actual causes (as beaming turns out to be only part of the explanation).

Here, we treat the photon directly, with a null stress-energy tensor, rather than as the $\delta\to0$ limit of a massive particle, which can be tricky. Three ingredients determine the result of our calculation. First, the photon is nearly resonant with the BH quasinormal modes. The source contains only the frequencies $\omega=m\Omega$, with $\Omega$ the orbital frequency of the photon and $m$ the azimuthal number, while in the eikonal limit the quasinormal frequencies are $m\Omega$ plus a small offset \cite{Cardoso:2008bp,Yang:2012he} and their damping rates are $(n+\tfrac12)\lambda_{\rm L}$, with $n$ the overtone number and $\lambda_{\rm L}$ the Lyapunov exponent of the light ring, the rate at which nearby null geodesics peel away from it. As we show in Secs.~\ref{sec:barrier} and \ref{sec:kerr}, the source sits within a few damping widths of resonance, independently of $\ell$ and of the spin; hence no multipole is exponentially suppressed.

Second, the two leading terms of the source, coming from the energy density $T_{tt}$ and the azimuthal pressure $T_{\phi\phi}$ of the particle, cancel exactly because $p^\mu p_\mu=0$. Without this cancellation the flux per multipole would grow like $\ell$; with it, it falls like $1/\ell$.

Third, the effective potential for gravitational perturbations is locally symmetric about its maximum (which in the eikonal limit coincides with the circular photon orbit), which makes the flux into the horizon equal to the flux to infinity at leading order.

For a massive particle at $r_0=3M(1+\delta)$ the first two ingredients fail beyond $\ell\sim\gamma^2\sim1/\delta$: the orbit is slightly outside the light ring, so the source frequencies $m\Omega$ drift away from the quasinormal frequencies as $m$ grows, and the rest mass spoils the cancellation. The loss of resonance dominates: the modes beyond $\ell\sim\gamma^2$ are no longer excited, and the spectrum is cut off exponentially.

Our analysis builds on tools developed in the '70s for the massive case. The perturbation equations are solved near the maximum of their effective potential, where the potential is well approximated quadratically; this gives the flux of the dominant, even-parity modes in closed form \cite{Breuer:1973kt,Breuer:1973bchm}. The same treatment yields the equality of the fluxes to infinity and into the horizon, derived there for scalar waves \cite{Breuer:1973bchm} and interpreted geometrically in Ref.~\cite{Hughes:1973ap}, and the dependence on the BH spin \cite{Chrzanowski:1974zb}. We make these connections explicit. What is new here is, first, the coefficient of the $1/\ell$ law for the odd-parity modes, which we show to be four times larger than in the 1973 formulae \cite{Breuer:1973kt,Ruffini:1973} (and in the amplitudes of Refs.~\cite{Breuer:1972cjr,Breuer:1975ncb}); second, the explicit null limit, which gives $\kappa$ in closed form and hence the proof of the divergence; third, the flux into the horizon for gravitational rather than scalar waves (anticipated by the geometric argument of Ref.~\cite{Hughes:1973ap}, which does not depend on the spin of the field), including the $O(\ell^{-1/2})$ asymmetry between the two fluxes; and fourth, the exact normalization in Kerr, where the WKB treatment of Ref.~\cite{Chrzanowski:1974zb} agrees with ours to $6\%$, partly by accident (Sec.~\ref{sec:kerr}).

The paper is organized as follows. Section~\ref{sec:setup} sets up the problem for a Schwarzschild BH, Sec.~\ref{sec:largel} contains the large-$\ell$ analysis and the main result, Sec.~\ref{sec:kerr} extends it to Kerr, Sec.~\ref{sec:1973} compares it with the formulae of the '70s, and Secs.~\ref{sec:disc} and \ref{sec:concl} discuss timelike orbits, the physical interpretation and possible applications. Throughout this paper we use geometrical units ($G=c=1$).

\section{The problem and the formalism}\label{sec:setup}

We start with the non-spinning case. Consider a massless particle of Killing energy $E$ on the circular null geodesic $r_0=3M$ of Schwarzschild, with angular momentum $L=bE$, impact parameter $b=3\sqrt3M$, and orbital frequency $\Omega=1/b$. With $\lambda$ an affine parameter and $p^\mu=\dd z^\mu/\dd\lambda$ its four-momentum, the stress-energy tensor of the particle is
\begin{align}
T^{\mu\nu}&=\int\dd\lambda\,\frac{p^\mu p^\nu}{\sqrt{-g}}\,\delta^4(x-z(\lambda))\nonumber\\
&=\frac{p^\mu p^\nu}{p^t r^2}\,\delta(r-r_0)\,\delta(\cos\theta)\,\delta(\phi-\Omega t),
\label{eq:T}
\end{align}
with $p^t=E/f(r_0)$, $f=1-2M/r$. This has the same form as the stress-energy tensor of a massive particle with four-velocity $u^\alpha$, $\mu u^\alpha u^\beta/(u^tr^2)$, with $\mu u^\alpha$ replaced by $p^\alpha$. We can therefore borrow the standard point-particle sources of the Zerilli and Regge--Wheeler equations \cite{Zerilli:1970se,Sago:2002fe,Martel:2005ir}, specialized to a circular orbit, replacing the energy and angular momentum per unit mass by $E_0\to E$, $L_0\to bE$. Every source coefficient is then proportional to $E$ at fixed $b$, so the fluxes scale as $E^2$. In the conventions of the \textsc{ReggeWheeler} package of the Black Hole Perturbation Toolkit \cite{BHPToolkit}, the stress-energy coefficients are
\begin{align}
\sA&=-\frac{16\pi(r_0-2M)\,E}{r_0^3}\,\bar Y_{\ell m},\nonumber\\
\sC&=-\frac{16\pi(r_0-2M)\,bE}{r_0^4\,\ell(\ell+1)}\,\partial_\theta\bar Y_{\ell m},\nonumber\\
\sF&=-\frac{16\pi(r_0-2M)\,b^2E}{r_0^5}\nonumber\\
&\qquad\times\frac{\ell(\ell+1)-2m^2}{(\ell-1)\ell(\ell+1)(\ell+2)}\,\bar Y_{\ell m},
\label{eq:ACF}
\end{align}
where $\bar Y_{\ell m}$ is the complex conjugate of the scalar spherical harmonic and all angular factors are evaluated at the particle, $(\theta,\phi)=(\pi/2,0)$.

The perturbation is decomposed into multipoles $(\ell,m)$, and for each of them a single master function $\Psi_{\ell m}$ obeys a wave equation in the tortoise coordinate $r_*$,
\begin{equation}
\big[\partial_{r_*}^2+\omega^2-V_\pm(r)\big]\Psi_{\ell m}=\mathcal S_{\ell m},\qquad\omega=m\Omega,
\label{eq:master}
\end{equation}
with the Zerilli potential $V_+$ when $\ell+m$ is even and the Regge--Wheeler potential $V_-$ when $\ell+m$ is odd. The potential is the barrier that a wave of multipole $\ell$ must cross to escape to infinity or to fall into the hole. For large $\ell$ it is dominated by the centrifugal term, $V\simeq\ell^2f/r^2$, whose maximum is at the light ring.

This is the connection between the light ring and the quasinormal modes: a quasinormal mode is a wave that leaks away from the top of this barrier in both directions, and in the eikonal limit its frequency is $\omega_{\rm QNM}=(\ell+\tfrac12)\Omega-i(n+\tfrac12)\lambda_{\rm L}$ \cite{Iyer:1986np,Cardoso:2008bp,Yang:2012he}, with $n$ the overtone number, $\Omega$ the orbital frequency of the light ring and $\lambda_{\rm L}$ its Lyapunov exponent, the rate at which nearby null geodesics peel away from it. Because the photon moves on a circle, each mode of the source oscillates at the single frequency $\omega=m\Omega$, which for $m$ close to $\ell$ is close to the real part of $\omega_{\rm QNM}$.

The source $\mathcal S_{\ell m}$ is concentrated on the orbit: it is a combination of $\delta(r-r_0)$ and $\delta'(r-r_0)$. Away from $r_0$ the master function therefore satisfies the homogeneous equation, and integrating Eq.~\eqref{eq:master} across $r_0$ shows that the $\delta$ term produces a discontinuity in $\partial_r\Psi$ and the $\delta'$ term a discontinuity in $\Psi$ itself. We denote these discontinuities, or jumps, by $[\Psi]\equiv\Psi(r_0^+)-\Psi(r_0^-)$ and $[\partial_r\Psi]\equiv\partial_r\Psi(r_0^+)-\partial_r\Psi(r_0^-)$; they are linear combinations of $\sA$, $\sC$, $\sF$ \cite{Sago:2002fe,BHPToolkit}, and they are all that is needed to specify the source.

The solution is built from two homogeneous solutions. $\Psi^{\rm in}$ is purely ingoing at the horizon, $\Psi^{\rm in}\to e^{-i\omega r_*}$ for $r_*\to-\infty$, and at infinity is a combination of incident and reflected waves, $\Psi^{\rm in}\to A^{\rm inc}e^{-i\omega r_*}+A^{\rm ref}e^{i\omega r_*}$; $\Psi^{\rm up}$ is purely outgoing at infinity, $\Psi^{\rm up}\to e^{i\omega r_*}$. Their Wronskian is $W=2i\omega A^{\rm inc}$.

The wave produced by the source is obtained from these two solutions by the usual Green's function construction; for a source confined to $r_0$ the result only involves the values and derivatives of $\Psi^{\rm in}$ and $\Psi^{\rm up}$ at $r_0$. The amplitude of the outgoing wave $e^{i\omega r_*}$ that reaches infinity is
\begin{equation}
Z^{\infty}=\frac{\Psi^{\rm in}[\partial_{r_*}\Psi]-[\Psi]\partial_{r_*}\Psi^{\rm in}}{W}\Big|_{r_0},
\label{eq:Z}
\end{equation}
and the amplitude $Z^{H}$ of the ingoing wave at the horizon is the same expression with $\Psi^{\rm up}$ in place of $\Psi^{\rm in}$. The Wronskian in the denominator makes these amplitudes independent of how the homogeneous solutions are normalized.

The energy fluxes are $\Edot^{\infty,H}_{\ell m}=c_\pm\,\omega^2|Z^{\infty,H}|^2$ with $c_+=(\ell-1)(\ell+2)/[4\pi\ell(\ell+1)]$ and $c_-=\ell(\ell+1)/[16\pi(\ell-1)(\ell+2)]$. These are single-$m$ fluxes; since the $m$ and $-m$ modes contribute equally, the physical flux in the multipole $\ell$ is twice the sum over $m>0$.

As a check of the source, we inserted the same null stress-energy tensor in the Teukolsky equation (spin weight $s=-2$, in the conventions of the Black Hole Perturbation Toolkit \cite{BHPToolkit}, with $\Upsilon_t=r_0^2p^t=27M^2E$) and recovered the Zerilli--Regge--Wheeler fluxes to the precision of the Black Hole Perturbation Toolkit's solutions, between nine and seventeen significant digits for $\ell\le5$.

\section{Fluxes from a photon on the Schwarzschild light ring}\label{sec:largel}

The calculation has three steps: expand the source at large $\ell$, construct the Green's function of the radial equation from the two homogeneous solutions near the top of the barrier, and apply it to the source. The physics is the following. The quasinormal modes are the poles of the Green's function in the complex frequency plane, and the source oscillates at $\omega=m\Omega$, close to the real part of the fundamental mode; the response is therefore nearly resonant, and is set by the distance of $\omega$ from the pole in units of the damping rate (Sec.~\ref{sec:barrier}). The source itself, however, is weaker than that of a massive particle, whose rest mass spoils a cancellation specific to null motion (Sec.~\ref{sec:source}). The wave is the convolution of the Green's function with the source; since the source is a combination of $\delta(r-r_0)$ and $\delta'(r-r_0)$, the convolution collapses to the Green's function and its derivative evaluated at $r_0$, weighted by the two jumps. This is the content of Eq.~\eqref{eq:Z}, which we evaluate in Sec.~\ref{sec:fluxes}.

\subsection{The source}\label{sec:source}
We set $m=\ell-j$, keep $j$ fixed and expand in $1/\ell$. The expansion is organized in this way because, as we shall see, the flux depends on $j$ only through coefficients that decrease rapidly with $j$, so that the modes with $m$ close to $\ell$ dominate. Inserting Eq.~\eqref{eq:ACF} with $r_0=3M$, $b^2=27M^2$ into the jump conditions, we find for even $j$ (Zerilli sector)
\begin{equation}
[\Psi]=\frac{8\pi E\,\bar Y_{\ell m}}{\ell(\ell+1)},\quad
[\partial_r\Psi]=-\frac{8\pi(2j+1)E}{\ell M}\,\bar Y_{\ell m}+O(\ell^{-2}),
\label{eq:jumps}
\end{equation}
and for odd $j$ (Regge--Wheeler sector)
\begin{equation*}
[\Psi]=-\frac{16\sqrt3\pi E\,\partial_\theta\bar Y_{\ell m}}{\ell(\ell+1)},\qquad
[\partial_r\Psi]=\frac{16\pi E\,\partial_\theta\bar Y_{\ell m}}{\sqrt3\,\ell(\ell+1)M}.
\end{equation*}
Since $\partial_{r_*}=f\partial_r$ and $f(3M)=1/3$, the tortoise-coordinate jumps that enter Eq.~\eqref{eq:Z} are $[\partial_{r_*}\Psi]=[\partial_r\Psi]/3$.

The quantity that matters most is the even-parity derivative jump, and the point is its $O(1/\ell)$ magnitude. To see why this is remarkable, recall how the source coefficients enter the jumps \cite{Sago:2002fe,BHPToolkit}: $\sC$ sources only the Regge--Wheeler sector, while the Zerilli jumps are built from $\sA$ and $\sF$, at leading order in $\ell$ as $[\Psi]\simeq-r_0^3\sA/[2\ell^2(r_0-2M)]$ and $[\partial_r\Psi]\simeq r_0^3\sA/[4(r_0-2M)^2]+\ell^2r_0^2\sF/[4(r_0-2M)]$. Naively the derivative jump is therefore $O(\ell^0)$: $\sA$, which comes from the energy density $T_{tt}$, is $O(\ell^0)$, and $\sF$, which comes from the azimuthal pressure $T_{\phi\phi}$, is $O(\ell^{-2})$ but enters multiplied by $\ell^2$. The two contributions, however, add up to
\begin{equation}
[\partial_r\Psi]_{O(\ell^0)}=\frac{4\pi\bar Y}{E}\Big(\frac{L^2}{r_0^3}-\frac{E^2}{r_0-2M}\Big)\propto p^\mu p_\mu,
\label{eq:cancel}
\end{equation}
which vanishes for a photon. The two leading contributions cancel exactly, and what is left is the $O(1/\ell)$ term of Eq.~\eqref{eq:jumps}. This is the cancellation announced in the Introduction. For a massive particle the $O(\ell^0)$ term equals $-4\pi\bar Y\mu^2/(Er_0)$; it is suppressed by $\gamma^{-2}=\mu^2/E^2$ but not by $\ell^{-1}$, and, as shown in Sec.~\ref{sec:disc}, it is what eventually cuts off the spectrum of timelike orbits.

\subsection{The barrier top}\label{sec:barrier}
We now turn to the homogeneous equation. Both potentials are $V=\ell^2f/r^2+O(\ell)$, with a maximum at $r=3M$, i.e.\ at the light ring: the waves emitted by the photon start from the top of their barrier. Expanding about the top in the tortoise coordinate, with $x=r_*-r_*(3M)$, $V=V_0-\tfrac12kx^2+\tfrac16V_3x^3+\dots$ with $V_0=[\ell(\ell+1)+O(1)]/(27M^2)$, $k=[2\ell^2+O(\ell)]/(729M^4)$, $V_3=[4\ell^2+O(\ell)]/(6561M^5)$; all the coefficients are $O(\ell^2)$, since the shape of the barrier does not depend on $\ell$. Near its maximum the barrier is thus an inverted parabola, which we refer to as the parabolic barrier.

The cubic term limits the range over which the quadratic approximation holds. Away from the top the solutions oscillate with phase $\int(\omega^2-V)^{1/2}\dd x$, and the cubic term shifts this phase by $\sim V_3x^3/\sqrt k$;\footnote{Writing $\omega^2-V=p_0^2+\delta$ with $p_0^2=\omega^2-V_0+\tfrac12kx^2$ and $\delta=-\tfrac16V_3x^3$, the WKB phase $\int(p_0^2+\delta)^{1/2}\dd x$ changes by $\int\delta/(2p_0)\,\dd x$; for $|x|\gg k^{-1/4}$ the quadratic term dominates, $p_0\simeq(k/2)^{1/2}x$, and the shift is $-(V_3/36)(2/k)^{1/2}x^3$. It is small for $|x|^3\ll\sqrt k/V_3\propto\ell^{-1}$, whereas the local condition $\delta\ll p_0^2$ only requires $|x|\ll k/V_3\sim M$.} the quadratic approximation for the potential is faithful only as long as this shift is small, i.e.\ for $|x|\ll(\sqrt k/V_3)^{1/3}\sim M\ell^{-1/3}$. We refer to this range as the parabolic region.

The next question is where the emission frequency sits with respect to the top. Since $\omega^2=(\ell-j)^2/(27M^2)$, its distance from the top, in units of the curvature of the barrier, is
\begin{align}
\epsilon&\equiv\omega^2-V_0=-\frac{(2j+1)\ell+O(1)}{27M^2},\nonumber\\
\eta_j&\equiv\frac{-\epsilon}{\sqrt{2k}}=j+\frac12+O(\ell^{-1}).
\label{eq:aj}
\end{align}
This dimensionless barrier parameter controls everything that follows, and it does not depend on $\ell$. In other words, the mode $(\ell,\ell-j)$ lies $(j+\tfrac12)\Omega$ below the real part $(\ell+\tfrac12)\Omega$ of the fundamental quasinormal frequency, i.e.\ $j+\tfrac12$ in units of the instability rate of the light ring, $\lambda_{\rm L}=\sqrt{2k}/(2\omega)=\Omega$ (for the parabolic barrier ${\rm Im}\,\omega_{\rm QNM}=-(n+\tfrac12)\sqrt{2k}/(2\omega)$, so this is the Lyapunov exponent of Sec.~\ref{sec:setup}, and it coincides with the orbital frequency). The source is thus at a distance $(j+\tfrac12)\lambda_{\rm L}$ from resonance, equal for $j=0$ to the damping rate of the fundamental mode, whatever the value of $\ell$.

Near the top the barrier is a parabola, and the radial equation can then be solved exactly. In the variable $z=(2k)^{1/4}x$ it becomes Weber's equation, $u_{zz}+(z^2/4-\eta)u=0$ (here and below $\eta=\eta_j$), the Schr\"odinger equation of an inverted harmonic oscillator; its solutions are the parabolic cylinder functions $D_\nu$ \cite{AbramowitzStegun,DLMF}. The quadratic approximation holds up to a cubic correction $\propto V_3z^3/(2k)^{5/4}=O(\ell^{-1/2})z^3$, which stays small throughout the parabolic region $|z|\ll\ell^{1/6}$.

We need the solution that is purely transmitted toward the horizon, $u=\mathcal C\,D_\nu(-e^{-i\pi/4}z)$ with $\nu=-\tfrac12-i\eta$. Its normalization is fixed by matching: at large $|z|$, but still inside the parabolic region ($1\ll|z|\ll\ell^{1/6}$), $u$ is a combination of incident and reflected WKB waves, and we choose $\mathcal C$ so that the incident wave has unit amplitude, $u=\Psi^{\rm in}/A^{\rm inc}$; the Wronskian $W=2i\omega A^{\rm inc}$ is thereby absorbed into $u$. By the connection formula for $D_\nu$ \cite{DLMF}, this gives $|\mathcal C|^2=\omega e^{-\pi\eta/2}/[(2k)^{1/4}\cosh\pi\eta]$ and reproduces the known transmission coefficient of a parabolic barrier \cite{Kemble:1935,LandauLifshitzQM,Iyer:1986np}, $|\mathcal T|^2=(1+e^{2\pi\eta})^{-1}$, which is $0.041$ for $\eta=\tfrac12$: the barrier reflects $96\%$ of the $m=\ell$ waves.

Since the source is a delta function at the top of the barrier, only the values of $u$ and of its derivative at $z=0$ are needed in what follows. With $D_\nu(0)=2^{\nu/2}\sqrt\pi/\Gamma(\tfrac{1-\nu}2)$, $D'_\nu(0)=-2^{(\nu+1)/2}\sqrt\pi/\Gamma(-\tfrac\nu2)$, and $\omega/(2k)^{1/4}=\sqrt{\ell/2}$, $\omega(2k)^{1/4}=\sqrt2\ell^{3/2}/(27M^2)$ at leading order, they are
\begin{align}
|u(0)|^2&=\sqrt{\frac\ell2}\,\frac{\pi\,e^{-\pi\eta/2}}{\sqrt2\cosh(\pi\eta)\,|\Gamma(\tfrac34+\tfrac{i\eta}2)|^2},\label{eq:u0}\\
|\partial_{r_*}u(0)|^2&=\frac{\sqrt2\,\ell^{3/2}}{27M^2}\,\frac{\sqrt2\,\pi\,e^{-\pi\eta/2}}{\cosh(\pi\eta)\,|\Gamma(\tfrac14+\tfrac{i\eta}2)|^2}.\label{eq:Du0}
\end{align}

The solution with unit incident amplitude from the horizon side, $v=\Psi^{\rm up}/A^{\rm inc}$ (the same $A^{\rm inc}$ enters $W$ on both sides), is obtained by reflecting the barrier: $v(z)=u(-z)$ up to the $O(\ell^{-1/2})$ cubic asymmetry, so $|v(0)|=|u(0)|$ and $\partial_{r_*}v(0)=-\partial_{r_*}u(0)$. This is the origin of the equality of the two fluxes: the wave is created at the top of the barrier, and from there the barrier looks the same in both directions.

\subsection{The fluxes}\label{sec:fluxes}
Combining Eq.~\eqref{eq:Z} with $W=2i\omega A^{\rm inc}$, the fluxes become $\Edot^{\infty,H}_{\ell m}=(c_\pm/4)\,|u(0)[\partial_{r_*}\Psi]\mp[\Psi]\partial_{r_*}u(0)|^2$. The two terms inside the modulus are the responses to the two parts of the source, the $\delta$ term (the derivative jump) and the $\delta'$ term (the jump in $\Psi$); the relative sign differs between the two fluxes because $\partial_{r_*}v(0)=-\partial_{r_*}u(0)$. Each factor carries a definite power of $\ell$. From Eq.~\eqref{eq:jumps}, $[\partial_{r_*}\Psi]\sim\ell^{-1}$ and $[\Psi]\sim\ell^{-2}$ times a harmonic; from Eqs.~\eqref{eq:u0}--\eqref{eq:Du0}, $|u(0)|^2\sim\ell^{1/2}$ and $|\partial_{r_*}u(0)|^2\sim\ell^{3/2}$; and the harmonics at the equator behave as $|Y_{\ell,\ell-j}(\pi/2)|^2\to c_j\sqrt\ell/(2\pi^{3/2})$, with $c_j=\binom{j}{j/2}2^{-j}$, and $|\partial_\theta Y_{\ell,\ell-j}|^2\to d_j\ell^{3/2}/\pi^{3/2}$, with $d_j=j\binom{j-1}{(j-1)/2}2^{1-j}$. In the even sector the dominant term is $|u(0)|^2[\partial_{r_*}\Psi]^2\sim\ell^{1/2}\ell^{-2}\ell^{1/2}=\ell^{-1}$; in the odd sector it is $|\partial_{r_*}u(0)|^2[\Psi]^2$, also $\propto\ell^{-1}$. The cross term is smaller by $\ell^{-1/2}$ and changes sign between the two fluxes.

Hence, for every $j$,
\begin{equation}
\Edot^{\infty,H}_{\ell,\ell-j}=\frac{\kappa_j}{\ell}\Big[1\pm\frac{\sigma_j}{\sqrt\ell}+O(\ell^{-1})\Big]\frac{E^2}{M^2},
\label{eq:result}
\end{equation}
with
\begin{align}
\kappa_j&=\frac{\sqrt\pi}{9}\,\frac{(2j+1)^2c_j\,e^{-\pi\eta_j/2}}{\cosh(\pi\eta_j)\,|\Gamma(\tfrac34+\tfrac{i}2\eta_j)|^2}\quad(j\ \text{even}),\label{eq:keven}\\
\kappa_j&=\frac{8\sqrt\pi}{9}\,\frac{d_j\,e^{-\pi\eta_j/2}}{\cosh(\pi\eta_j)\,|\Gamma(\tfrac14+\tfrac{i}2\eta_j)|^2}\quad(j\ \text{odd}),\label{eq:kodd}
\end{align}
$\eta_j=j+\tfrac12$. The whole $\eta$ dependence of the even coefficients sits in the function $F(\eta)\equiv e^{-\pi\eta/2}/[\cosh(\pi\eta)|\Gamma(\tfrac34+\tfrac{i\eta}2)|^2]$, so that $\kappa_j=(\sqrt\pi/9)(2j+1)^2c_jF(\eta_j)$ for even $j$; $F$ will reappear in Secs.~\ref{sec:kerr} and \ref{sec:disc}.

Numerically $\kappa_0=0.0277668$, $\kappa_1=0.0037670$, $\kappa_2=2.7521\times10^{-4}$, $\kappa_3=1.6604\times10^{-5}$, $\kappa_4=9.23\times10^{-7}$. The $\kappa_j$ fall off roughly like $e^{-\pi j}$ (up to powers of $j$), so the sum over $m$ converges quickly, $\sum_j\kappa_j=0.0318265$, and including the $-m$ modes
\begin{equation}
\Edot_\ell^{\infty}=\Edot_\ell^{H}\to\frac{\kappa E^2}{M^2\ell},\qquad\kappa=2\sum_j\kappa_j=0.063653.
\label{eq:kappa}
\end{equation}
This is the main result for Schwarzschild: the sum over $\ell$ diverges logarithmically, and the coefficient agrees with the fit $\kappa=0.064\pm0.001$ of Ref.~\cite{Barausse:2021ryv}.

Two remarks are in order. The first concerns the equality $\Edot^{\infty}=\Edot^{H}$. It does not mean that half of the wave tunnels through the barrier: the barrier transmits only $4\%$ of the $m=\ell$ waves. It means that the wave is created at the top of the barrier, and that from the top the barrier looks the same toward the horizon and toward infinity, $|u(0)|=|v(0)|$, so that the two waves launched in the two directions carry the same energy. The statement holds mode by mode, not only after summing over $m$.

The second remark concerns how the two fluxes approach their common limit, i.e.\ the coefficient $\sigma_j$ of Eq.~\eqref{eq:result}, which enters the two fluxes with opposite signs. Because of the opposite signs the $O(\ell^{-1/2})$ correction cancels in the sum of the two fluxes, which approaches its limit with relative corrections $O(1/\ell)$, but not in their difference.

The coefficient $\sigma_j$ has two sources. The first is the cross term in the flux formula above, whose value follows from the phase of
\begin{equation}
\frac{u(0)}{\partial_{r_*}u(0)}=\frac{e^{i\pi/4}\,\Gamma(\tfrac14+\tfrac{i\eta}2)}{\sqrt2\,(2k)^{1/4}\,\Gamma(\tfrac34+\tfrac{i\eta}2)}.
\label{eq:phase}
\end{equation}
The second is the cubic term of the barrier, which makes the two sides slightly different at this order, $|v(0)|\neq|u(0)|$. For $j=0$ the cross term, which follows from Eq.~\eqref{eq:phase}, contributes $+1.039$, and the cubic term, which we evaluate by solving Weber's equation with the cubic correction $O(\ell^{-1/2})z^3$ included and extracting the $O(\ell^{-1/2})$ change of $|v(0)|^2/|u(0)|^2$, contributes $-0.23$, so that $\sigma_0=0.81$; the $\sigma_j$ with $j\ge1$, obtained numerically, are negative, and after summing over $m$ the relative difference between the two fluxes is
\begin{equation*}
\frac{\Edot^\infty_\ell-\Edot^H_\ell}{\Edot^\infty_\ell+\Edot^H_\ell}\to\frac{\bar\sigma}{\sqrt\ell},\qquad\bar\sigma=0.61.
\end{equation*}

Two consequences follow. At $\ell\simeq100$ the two fluxes still differ by $2\bar\sigma/\sqrt\ell\simeq12\%$, as is visible in the numerical data below: this is the rate at which the equality $\Edot^\infty=\Edot^H$, reported in Ref.~\cite{Barausse:2021ryv} on the basis of the fitted $\kappa$, is approached. If the fluxes are summed over $\ell$ up to some $\ell_{\max}$ and one asks what fraction of the total has gone into the horizon, the $\ell^{-1/2}$ terms do not drop out, because they cancel in the sum of the two fluxes but not in their ratio. The fraction therefore approaches $1/2$ only logarithmically: with the asymptotic formulae it is $40\%$ for $\ell_{\max}=100$--$140$. This is why Ref.~\cite{Gundlach:2012aj}, which truncates its mode sums at $\ell\simeq140$ and extrapolates in radius to the light ring, found $42\%$ rather than $50\%$.

\subsection{Numerical check}\label{sec:numerics}
To check these results we solved the problem numerically, without any large-$\ell$ approximation. For each $(\ell,m)$ we integrated the homogeneous Zerilli or Regge--Wheeler equation in $r_*$ with a ninth-order explicit Runge--Kutta method, twice: from $r_*=-40M$, starting from a purely ingoing wave, to obtain $\Psi^{\rm in}$, and from $r=40M$, starting from a purely outgoing wave, to obtain $\Psi^{\rm up}$. The two solutions and their derivatives at $r_0$, together with the jumps of Sec.~\ref{sec:source}, give the amplitudes of Eq.~\eqref{eq:Z} and hence the fluxes: this is the Green's function construction of Sec.~\ref{sec:setup}, carried out exactly rather than near the top of the barrier. The integrator was validated against the solutions of the Black Hole Perturbation Toolkit \cite{BHPToolkit} at $\ell\le5$, where the two agree to $10^{-7}$ at $\ell=5$ ($10^{-5}$ at $\ell=2$), and was then pushed to $\ell=12800$. The integrator, the Kerr code of Sec.~\ref{sec:kerr}, the scripts that produce the figures and the numerical data are publicly available \cite{notebook}.

\begin{figure}
\includegraphics[width=\columnwidth]{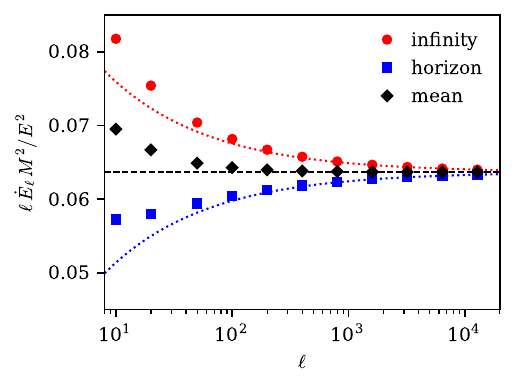}
\caption{Flux per multipole from a photon on the Schwarzschild light ring, summed over the five dominant $m$ ($j\le4$) and multiplied by $\ell$, from the numerical solutions: to infinity (circles), into the horizon (squares) and their mean (diamonds). The dashed line is the analytic coefficient $\kappa=0.063653$ of Eq.~\eqref{eq:kappa}; the dotted lines are $\kappa(1\pm\bar\sigma/\sqrt\ell)$, the two fluxes with their leading $O(\ell^{-1/2})$ corrections.}
\label{fig:schw}
\end{figure}

Figure~\ref{fig:schw} shows the result. The symbols are the numerical fluxes per multipole, summed over the five dominant $m$ ($j\le4$, which captures all but $2\times10^{-6}$ of the sum) and multiplied by $\ell$, so that a $1/\ell$ law appears as a constant: circles for the flux to infinity, squares for the flux into the horizon, diamonds for their mean. The dashed horizontal line is the analytic coefficient $\kappa=0.063653$ of Eq.~\eqref{eq:kappa}. The two dotted curves are $\kappa(1\pm\bar\sigma/\sqrt\ell)$, i.e.\ the two fluxes with the leading correction of Eq.~\eqref{eq:result} included. The flux to infinity approaches $\kappa$ from above and the flux into the horizon from below, both following the dotted curves, while their mean sits on the dashed line already at $\ell\sim100$.

Quantitatively, a three-term fit of the mean flux, $a_0+a_2/\ell+a_3\ell^{-3/2}$ (the $\ell^{-1/2}$ term cancels in the mean), over $400\le\ell\le12800$ gives $\sum_j\kappa_j=0.0318263$ for the single-$m$ sum, against the analytic $0.0318265$. At $\ell=12800$ the individual $\kappa_j$, $j\le4$, are reproduced to better than $10^{-3}$ in relative terms, and $\kappa_0$ and $\kappa_1$ to about $10^{-4}$ and $2\times10^{-5}$. Finally, since the numerical solutions give $u(0)$, $\partial_{r_*}u(0)$, $v(0)$ and $\partial_{r_*}v(0)$ separately, the two contributions to $\sigma_0$ can be isolated: at $\ell=6400$, the largest $\ell$ for which we computed $v$ as well, the cross term is $1.0389$, against the analytic $1.0390$, and the barrier asymmetry is $-0.231$.

\section{Extension to Kerr black holes}\label{sec:kerr}
\begin{table*}
\caption{Results for circular photon orbits in Kerr (negative $a$ denotes retrograde orbits): radius, impact parameter, $\hat\Upsilon_t$, barrier curvature $\tilde k$, source coefficient $|\Ah_0|$, amplitude factor $\mathcal N$, the single-$m$ coefficient $\kappa_0(a)$ of Eq.~\eqref{eq:kappakerr} and the spin factor $g(a)$. Last column: ratio of the numerical mean flux $\tfrac12(\Edot^\infty_\ell+\Edot^H_\ell)$ at $\ell=800$, summed over the four dominant $|m|$ ($j\le3$, both signs of $m$) and multiplied by $\ell$, to the analytic value with the same truncation, $2g(a)\sum_{j\le3}\kappa_j$. No numerical runs were made at $|a|/M=0.99$; for the prograde orbit the barrier is in any case so flat that the asymptotic regime sets in only at $\ell\gg800$.}
\label{tab:kerr}
\begin{ruledtabular}
\begin{tabular}{rccccccccc}
$a/M$ & $r_0/M$ & $b/M$ & $\hat\Upsilon_t/M^2$ & $M^2\tilde k$ & $|\Ah_0|/M^2$ & $M^4\mathcal N$ & $\kappa_0(a)$ & $g(a)$ & num./an.\\
\hline
$0$ & 3 & 5.196 & 27 & $7.407\times10^{-2}$ & 11.69 & $2.195\times10^{-2}$ & 0.02777 & 1 & 1.0012\\
$0.5$ & 2.347 & 4.096 & 21.87 & $3.408\times10^{-2}$ & 5.477 & $9.566\times10^{-2}$ & 0.03428 & 1.235 & 1.0014\\
$0.9$ & 1.558 & 2.844 & 19.83 & $1.948\times10^{-3}$ & 1.418 & 1.119 & 0.03781 & 1.362 & 1.0017\\
$0.99$ & 1.168 & 2.252 & 33.89 & $1.825\times10^{-5}$ & 0.5258 & 6.313 & 0.02212 & 0.797 & --\\
$-0.5$ & 3.532 & 6.138 & 32.18 & $9.149\times10^{-2}$ & 18.89 & $8.240\times10^{-3}$ & 0.02329 & 0.839 & 1.0011\\
$-0.9$ & 3.910 & 6.832 & 36.31 & $9.360\times10^{-2}$ & 25.23 & $4.476\times10^{-3}$ & 0.02065 & 0.744 & 1.0010\\
$-0.99$ & 3.991 & 6.983 & 37.23 & $9.317\times10^{-2}$ & 26.72 & $3.959\times10^{-3}$ & 0.02014 & 0.725 & --\\
\end{tabular}
\end{ruledtabular}
\end{table*}

For a photon on the equatorial light ring of Kerr, we work directly with the Teukolsky equation for spin weight $s=-2$ \cite{Teukolsky:1973ha}, for a BH of mass $M$ and spin $a$, in Boyer--Lindquist coordinates. The logic is the same as in Schwarzschild: we locate the barrier top, compute the barrier parameter $\eta_j$, build the Green's function from the two homogeneous solutions, and convolve it with the source. The new elements are a complex potential, spheroidal rather than spherical harmonics, and a source whose leading terms must again be shown to cancel.

We recall that the Teukolsky equation describes the perturbation through the Weyl scalar $\psi_4$, whose radial part $R(r)$ obeys a second-order equation and whose angular part is a spin-weighted spheroidal harmonic $S(\theta)$ with separation constant $\Lambda$ \cite{Teukolsky:1973ha}. The fluxes are read off from the asymptotic amplitudes of $R$, with a known factor (the Teukolsky--Press factor \cite{Teukolsky:1974yv}) converting the amplitude at the horizon into an energy flux. Retrograde orbits are obtained from prograde ones by $a\to-a$.

The geometry of the light ring is fixed by the radial potential of equatorial null geodesics \cite{Bardeen:1972fi}, $\tilde R=P^2-\Delta(b-a)^2$ with $\Delta=r^2-2Mr+a^2$ and $P=r^2+a^2-ab$: the radius $r_0(a)=2M\{1+\cos[\tfrac23\arccos(-a/M)]\}$ \cite{Bardeen:1972fi} and the impact parameter $b$ follow from $\tilde R(r_0)=\tilde R'(r_0)=0$ (primes denote $r$ derivatives, and for $P$ and $\Delta$ a subscript $0$ denotes evaluation at $r_0$). It is convenient to use $r_0$ itself as the parameter: the light-ring equation $r_0(r_0-3M)^2=4Ma^2$ gives the spin for which the light ring sits at radius $r_0$, $a=\sqrt{r_0/M}\,(3M-r_0)/2$, with $M<r_0<4M$ (prograde orbits for $r_0<3M$), and then
\begin{equation}
b=\frac{\sqrt{r_0/M}\,(r_0+3M)}{2},\qquad
\tilde R=(r-r_0)^2\,r\,(r+2r_0).
\label{eq:closed}
\end{equation}
The orbital frequency is $\Omega=1/b$, and the time component of the momentum enters through $\Upsilon_t\equiv\Sigma\,\dd t/\dd\lambda=E\hat\Upsilon_t$ ($\lambda$ the affine parameter), with $\hat\Upsilon_t=(r_0^2+a^2)P_0/\Delta_0+a(b-a)=r_0^2(r_0+3M)/(r_0-M)$ and $\Sigma=r^2+a^2\cos^2\theta$.

The meaning of $\tilde R(r_0)=0$ is worth spelling out. For a particle of mass $\mu$ with the same $E$, $L$ and $r_0$, the radial geodesic equation reads $\Sigma^2(\dd r/\dd\lambda)^2=E^2\tilde R-\mu^2r^2\Delta$, so on a circular orbit $\tilde R(r_0)=0$ is the condition $\mu=0$: it is the Kerr version of the null condition $L^2/r_0^3=E^2/(r_0-2M)$ that produced the cancellation of Eq.~\eqref{eq:cancel}.

The projection of the source onto the Teukolsky equation is the standard circular-orbit source term $\alpha_{\ell m}$ of Ref.~\cite{Hughes:1999bq}, which we use in the conventions of the \textsc{Teukolsky} package of the Black Hole Perturbation Toolkit \cite{BHPToolkit}. It is a combination of $R$, $R'$ and $R''$ at $r_0$ and of the spheroidal harmonic $S$ and its derivatives at $\theta=\pi/2$. The null geodesic enters it only through $E$, $L=bE$ and $\Upsilon_t$, so that $\alpha_{\ell m}\propto E^2$. The flux at infinity is $|Z^{\infty}|^2/(4\pi\omega^2)$ with $Z^{\infty}=-8\pi\alpha^{\rm in}/(W\Upsilon_t)$, where $\alpha^{\rm in}$ is $\alpha_{\ell m}$ evaluated on the solution $R^{\rm in}$ that is ingoing at the horizon (the analogue of $\Psi^{\rm in}$), $W=2i\omega B^{\rm inc}$, and $B^{\rm inc}$ is the amplitude of the incident wave in $R^{\rm in}$ at infinity \cite{Hughes:1999bq}; the horizon flux is $|Z^H|^2/(4\pi\omega^2)$ with $Z^H$ built from the solution $R^{\rm up}$ that is outgoing at infinity and multiplied by the Teukolsky--Press factor \cite{Teukolsky:1974yv}.

The Teukolsky equation does not have the Schr\"odinger form of the Zerilli equation, but it can be brought to it \cite{Teukolsky:1974yv}. With $R=\Delta^{-s/2}(r^2+a^2)^{-1/2}\mathcal Y$ the radial equation becomes $\mathcal Y_{,r_*r_*}+Q\mathcal Y=0$ with the complex potential
\begin{equation}
Q=\frac{K^2-2is(r-M)K+\Delta(4is\omega r-\Lambda)}{(r^2+a^2)^2}-G^2-\frac{\dd G}{\dd r_*},
\end{equation}
where $G=s(r-M)/(r^2+a^2)+r\Delta/(r^2+a^2)^2$, $K=(r^2+a^2)\omega-am$, and $\Lambda$ is the spheroidal separation constant.

The imaginary part of $Q$ is the new feature. For the resonant modes, $m=\omega b$, it is ${\rm Im}\,Q=s\omega[-2(r-M)P+4r\Delta]/(r^2+a^2)^2$, and the two light-ring conditions give $-2(r_0-M)P_0+4r_0\Delta_0=0$. So ${\rm Im}\,Q$ vanishes exactly at the light ring, for every spin, and is odd in $x=r_*-r_*(r_0)$ at leading order. Since ${\rm Im}\,Q=O(\omega x/M^2)$ while the curvature term of ${\rm Re}\,Q$ is $O(\omega^2x^2/M^4)$, in the variable $z$ of Weber's equation it is an $O(\ell^{-1/2})z$ odd perturbation, like the cubic term of the barrier, and it affects neither the leading order nor the sum of the two fluxes. Away from the light ring it is responsible for the familiar $r^{\pm s}$ and $\Delta^{\pm s/2}$ behaviors of the Teukolsky function \cite{Teukolsky:1973ha}, which we take into account through the WKB amplitude integrals defined below.

The real part of $Q$ requires the separation constant $\Lambda$. For $m=\ell-j$ and $a\omega=O(\ell)$ it follows from expanding the angular equation about the equator, which turns it into the equation of a harmonic oscillator whose $j$-th eigenvalue gives \cite{Chrzanowski:1974zb,Yang:2012he} $\Lambda=(m-a\omega)^2+(2j+1)\omega\beta_b+O(1)$, with $\beta_b=\sqrt{b^2-a^2}=\sqrt3\,r_0$ (we checked this against the eigenvalues computed by the Black Hole Perturbation Toolkit \cite{BHPToolkit}).

With it, ${\rm Re}\,Q=\omega^2\tilde R/(r^2+a^2)^2-(2j+1)\omega\beta_b\Delta/(r^2+a^2)^2+O(1/M^2)$, so the barrier top is again at the light ring, with curvature $k=\omega^2\tilde k$, $\tilde k=\Delta_0^2\tilde R''(r_0)/(r_0^2+a^2)^4=3r_0^4(r_0-M)^4/[8M^2(r_0^2+a^2)^4]$, and the barrier parameter is
\begin{equation}
\eta_j=\Big(j+\frac12\Big)\frac{\Omega_\theta}{\lambda_{\rm L}},\quad
\Omega_\theta=\frac{\beta_b}{\hat\Upsilon_t},\quad
\lambda_{\rm L}=\sqrt{\frac{\tilde R''(r_0)}{2\hat\Upsilon_t^2}},
\end{equation}
where $\Omega_\theta$ is the vertical epicyclic frequency and $\lambda_{\rm L}$ the Lyapunov exponent of the light ring \cite{Cardoso:2008bp}.

The two are equal: from Eq.~\eqref{eq:closed}, $\tilde R''(r_0)=6r_0^2$ and $b^2-a^2=3r_0^2$, hence $\Omega_\theta=\lambda_{\rm L}$ for every Kerr equatorial light ring. So $\eta_j=j+\tfrac12$ independently of the spin, and the radial solutions at the barrier top are the same as in Schwarzschild.

As for the source, we expand $\alpha_{\ell m}$ in powers of $\omega$, treating the value of the wave at the light ring, $\mathcal Y(0)$, and its derivative, $\mathcal Y_{,r_*}(0)=\omega^{1/2}y_1$, as unknowns. The leading terms, of order $\omega^2$ and $\omega^{3/2}$ for even $j$ and of order $\omega$ for odd $j$, cancel for a null source; both cancellations are consequences of the null condition $\tilde R(r_0)=0$ discussed above, and we verified them algebraically for every spin, using $r_0$ as the parameter. What remains is $\alpha=E^2\Ah_j\,\omega\,S_0\,\mathcal Y(0)+O(\omega^{1/2})$ for even $j$, with $S_0=S(\pi/2)$ and $\Ah_j=(2j+1)\Ah_0$, and $\alpha=E^2\Bh\,\omega^{1/2}\partial_\theta S_0\,y_1+O(\omega^0)$ for odd $j$, with $\Bh$ independent of $j$; in Schwarzschild $\Ah_0=\tfrac94\sqrt{27}\,M^2=\tfrac34r_0b$ and $\Bh=-i\Ah_0$. The coefficients come out as elementary functions of $r_0$, $\Ah_0=\sqrt3\,r_0^4/(4M\sqrt{r_0^2+a^2})$ and $\Bh=-i\,r_0^{5/2}\sqrt{r_0^2+a^2}/[2\sqrt M\,(r_0-M)]$; their derivation, like all the calculations of this paper, can be reproduced with the companion Mathematica notebook and the SymPy script of Ref.~\cite{notebook}.

The last ingredient is the amplitude of the homogeneous solution at the light ring. In Eq.~\eqref{eq:Z} the wave at the light ring is divided by the Wronskian, i.e.\ by its incident amplitude at infinity, so what is needed is the ratio of the two, $u(0)$ in the notation of Sec.~\ref{sec:barrier}. In Schwarzschild the matching of that section gave it directly, because the WKB wave coming from infinity reaches the parabolic region with its incident amplitude unchanged. In Kerr this is no longer true.

With a complex potential the WKB solutions are $Q^{-1/4}\exp(\pm i\int\sqrt Q\,\dd r_*)$ with $\sqrt Q\simeq p+iq$, $p=\sqrt{{\rm Re}\,Q}$, $q={\rm Im}\,Q/(2p)$ (the local wavenumber and damping, not to be confused with the momentum $p^\mu$); $q$ is $\omega$ independent at leading order and tends to $s/r$ at infinity. Because $Q$ is complex, the WKB amplitude of a wave is not conserved as it propagates: the imaginary part of $\sqrt Q$ acts as a position-dependent gain or loss. This is not a physical absorption but a property of the field variable (the Teukolsky function $R$, which is $\psi_4$ rescaled by $(r-ia\cos\theta)^4$, grows like $r^3$ at infinity), and it has to be tracked to relate the amplitude at infinity, which defines the flux, to the amplitude at the light ring, which the source sees. The integrals $J_\infty$ and $h_\infty$ below do exactly that. The incident WKB wave then reaches the matching region with $|\mathcal Y_{\rm inc}|^2=|B^{\rm inc}|^2(\omega/p)\,\mathcal N$, where $\omega/p$ is the usual WKB amplitude factor $|Q|^{-1/2}$ referred to infinity, where $p\to\omega$, and the amplitude factor $\mathcal N$ is
\begin{align}
\mathcal N&=r_0^{2s}\,e^{-2sh_\infty}\,e^{-2J_\infty},\qquad
J_\infty=\int_{r_0}^\infty\Big(q-\frac sr\Big)\dd r_*,\nonumber\\
h_\infty&=\int_{r_0}^\infty\Big[\frac{r^2+a^2}{r\Delta}-\frac1r\Big]\dd r,\nonumber\\
q\,\dd r_*&=\frac{s[-2(r-M)P+4r\Delta]}{2\Delta\sqrt{\tilde R}}\,\dd r.
\label{eq:N}
\end{align}
Because $\tilde R$ has a double root at $r_0$, these integrals are elementary logarithms, and their exponentials collapse to the rational $\mathcal N=16M^2/r_0^6$. The matching of Sec.~\ref{sec:largel} gives $|u(0)|^2=\mathcal N(\omega/(2k)^{1/4})(\pi/\sqrt2)F(\eta_j)$, with $F(\eta)$ the function defined after Eq.~\eqref{eq:kodd} and $u=\mathcal Y^{\rm in}/B^{\rm inc}$. With $S$ normalized to unity on the sphere, the result for the dominant even modes is
\begin{equation}
\kappa_j(a)=\frac{\sqrt{2\pi}\,M^2b\,c_j\sqrt{\beta_b}\,\mathcal N\,F(\eta_j)\,|\Ah_j|^2}{(2\tilde k)^{1/4}\,\hat\Upsilon_t^2}\quad(j\ \text{even}),
\label{eq:kappakerr}
\end{equation}
where $c_j$ is the harmonic coefficient of Sec.~\ref{sec:fluxes}: the oscillator expansion of the angular equation used above for $\Lambda$ gives, at the equator and at large $\ell$, $|S(\pi/2)|^2\to c_j\sqrt{\omega\beta_b}/(2\pi^{3/2})$, which reduces to the spherical result of Sec.~\ref{sec:fluxes} at $a=0$, where $\omega\beta_b=m\simeq\ell$ (we checked this against the Toolkit's harmonics); the odd modes are obtained in the same way from $|\partial_{r_*}u(0)|^2$ and $\Bh$. For $a=0$, Eq.~\eqref{eq:kappakerr} and its odd counterpart reproduce all the Zerilli--Regge--Wheeler $\kappa_j$ to the digits quoted above, which checks the whole calculation.

Finally, since $\eta_j$, $c_j$, $d_j$ and $\Ah_j/\Ah_0$ do not depend on the spin, and $|\Bh/\Ah_0|^2\sqrt{2\tilde k}\,\beta_b=2$ for every spin (an identity that follows from the closed forms above), the ratios $\kappa_j(a)/\kappa_0(a)$ are spin independent and the whole spin dependence reduces to one factor,
\begin{equation}
\Edot^{\infty}_\ell=\Edot^{H}_\ell\to g(a)\,\frac{\kappa E^2}{M^2\ell},\qquad g(a)=\frac{\kappa_0(a)}{\kappa_0(0)}.
\label{eq:g}
\end{equation}
The flux into the horizon is obtained in the same way from the horizon side. There the normalization is $W=2iC^{\rm out}(r_+^2+a^2)(\varpi-2is\epsilon_+)$, with $\varpi=\omega-m\Omega_H$, $r_+=M+\sqrt{M^2-a^2}$ the horizon radius, $\Omega_H=a/(2Mr_+)$ its angular velocity, $\epsilon_+=\sqrt{M^2-a^2}/(4Mr_+)$ and $C^{\rm out}$ the amplitude of $R^{\rm up}$ at the horizon, the amplitude integral is $J_H=\int_{r_+}^{r_0}[q+s(r-M)/(r^2+a^2)]\dd r_*$, and the Teukolsky--Press factor converts $|Z^H|^2$ into a flux. The ratio of the horizon to the infinity flux built from these factors is exactly one for all spins,\footnote{We first found this numerically, to better than $10^{-17}$ at all the spins of Table~\ref{tab:kerr}, and then proved it with the closed forms above and with $J_H$, which, like $J_\infty$ and $h_\infty$, reduces to logarithms because of the double root of $\tilde R$: $e^{2J_H}=[r_0^2+Mr_0+4M^{3/2}\sqrt{4M-r_0}-8M^2]^4/r_0^8$.} so $\Edot^H=\Edot^\infty$ at leading order also in Kerr. Unlike in Schwarzschild, we have not computed the $O(\ell^{-1/2})$ difference between the two fluxes analytically for $a\neq0$; numerically, $\sqrt\ell(\Edot^\infty_\ell-\Edot^H_\ell)/(\Edot^\infty_\ell+\Edot^H_\ell)$ at $\ell=800$ is $0.78$, $1.10$, $0.51$ and $0.45$ for $a/M=0.5$, $0.9$, $-0.5$ and $-0.9$, against $0.61$ at $a=0$ (the asymptotic value of Sec.~\ref{sec:fluxes}, which the $a=0$ numerics reach by $\ell=800$). Note that $\Omega>\Omega_H$ for every light ring, so there is no superradiance for these modes.

With all the ingredients of Eq.~\eqref{eq:kappakerr} in closed form, the spin factor of Eq.~\eqref{eq:g} is
\begin{equation}
g(a)=\frac{27M^2(r_0-M)}{r_0^2(r_0+3M)}.
\label{eq:gclosed}
\end{equation}
It is maximal at $r_0=\sqrt3M$, i.e.\ at $a/M=3^{1/4}(3-\sqrt3)/2\simeq0.834$, where $g=6\sqrt3-9\simeq1.392$; it vanishes as $r_0\to M$ in the extremal prograde limit, and equals $81/112\simeq0.723$ in the extremal retrograde limit, $r_0=4M$.

Table~\ref{tab:kerr} lists the ingredients of Eq.~\eqref{eq:kappakerr}, as evaluated numerically by the companion Mathematica notebook, and the resulting $g(a)$; they agree with the closed forms and with Eq.~\eqref{eq:gclosed} to the working precision. For prograde orbits $g(a)$ grows with the spin up to its maximum near $a/M=0.83$ and then drops, to $0.99$ at $a/M=0.98$ and $0.62$ at $a/M=0.995$, as the Boyer--Lindquist radius of the light ring approaches that of the horizon \cite{Bardeen:1972fi} and the barrier flattens ($\tilde k\to0$, so that the asymptotic regime recedes to very large $\ell$).

This is the behavior found in Ref.~\cite{Chrzanowski:1974zb} for the $j=0$ term of massive particles: in the null limit, their Eq.~(4.28), with their Eq.~(2.35b) for the cut-off harmonic $m_{\rm crit}$ and $\mu^2\gamma^2=E^2$, gives a power to infinity $(12/\sqrt\pi)\,e^{-\pi/2}E^2(r_0-M)/[r_0^2(r_0+3M)m]$ at the frequency $m\Omega$, whose spin dependence coincides exactly with Eq.~\eqref{eq:gclosed}. Their power at a given frequency collects the modes $m$ and $-m$ (we checked this on the scalar formula of Ref.~\cite{Breuer:1973bchm}, which their master formula reproduces, against an exact scalar flux), so it has to be compared with $2\kappa_0g(a)E^2/(M^2m)$: their constant is $0.94$ times ours. The agreement is partly an accident, in which two errors of their WKB treatment compensate. In the massive regime, where their WKB approximation is valid, their scalar formula agrees with the exact scalar result of Ref.~\cite{Breuer:1973bchm}, but their tensor formula tends to $4$ times the exact result of Ref.~\cite{Breuer:1973kt} (Sec.~\ref{sec:1973}), once the two are referred to the same counting of $m$: the factor is the $(s!)^2$ of their master formula, Eq.~(5.1). In the null limit it is compensated by the WKB form of the barrier-top factor, $\eta^{3/2}e^{-\pi\eta}$, the large-$\eta$ form of the exact $(\eta+\tfrac12)^2F(\eta)$ obtained with Stirling's approximation of the Gamma function, which at $\eta=\tfrac12$ underestimates the exact one by $4\times1.065$; what remains is the factor $1/1.065=0.94$ (this factor $4$ is unrelated to the odd-parity factor $4$ of Sec.~\ref{sec:1973}).

\begin{figure}
\includegraphics[width=\columnwidth]{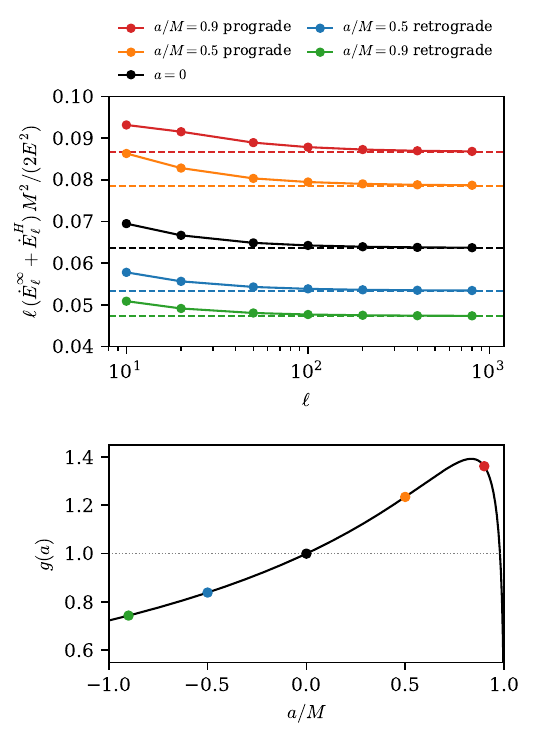}
\caption{Top: mean of the fluxes to infinity and into the horizon per multipole, summed over the four dominant $|m|$ ($j\le3$, both signs of $m$) and multiplied by $\ell$, for a photon on the light ring of Kerr black holes, from numerical solutions of the Teukolsky equation (symbols) and the analytic values with the same truncation, $2g(a)\sum_{j\le3}\kappa_j$ (dashed). Bottom: the spin factor $g(a)$ of Eq.~\eqref{eq:gclosed}, with the spins of the top panel marked; negative $a$ denotes retrograde orbits.}
\label{fig:kerr}
\end{figure}

Numerical solutions of the Teukolsky equation up to $\ell=800$ (eighth-order Dormand--Prince integrator), summed over the four dominant $m$ ($j\le3$; Fig.~\ref{fig:kerr}), confirm the analytic values: a Richardson extrapolation in $1/\ell$ from $\ell=400$ and $800$ agrees with $2g(a)\sum_{j\le3}\kappa_j$ to better than $1.5\times10^{-5}$ in relative terms in all five cases (the omitted $j\ge4$ terms contribute $3\times10^{-5}$ of $\kappa$).

\section{Comparison with the geodesic synchrotron radiation literature}\label{sec:1973}

Breuer, Ruffini, Tiomno and Vishveshwara \cite{Breuer:1973kt} derived, for a massive particle at $r_0=3M(1+\delta)$ and $|m|\gg1$, the power in the dominant even mode ($\ell=m$) and in the dominant odd mode ($\ell=m+1$). In the notation of this paper (their $\delta$ is three times ours, and their $\epsilon_{\rm even}=2\eta$ and $\epsilon_{\rm odd}=2\eta'$, with $\eta=\tfrac12+\tfrac32m\delta$ and $\eta'=\eta+1$), their Eqs.~(14)--(15) read
\begin{align}
\mathcal P^{\rm even}&=\frac{\mu^2}{M^2}\frac{e^{-\pi\eta/2}}{162\pi^{3/2}m\delta}\Big|(\eta+\tfrac12)\Gamma(\tfrac14+\tfrac i2\eta)\nonumber\\
&\qquad+\tfrac{\sqrt2(1-i)}{\sqrt{3m}}\Gamma(\tfrac34+\tfrac i2\eta)\Big|^2,\label{eq:brtv14}\\
\mathcal P^{\rm odd}&=\frac{\mu^2}{M^2}\frac{e^{-\pi\eta'/2}}{162\pi^{3/2}m\delta}\Big|\sqrt2\,\Gamma(\tfrac34+\tfrac i2\eta')\nonumber\\
&\qquad+\tfrac{1+i}{2\sqrt{3m}}\Gamma(\tfrac14+\tfrac i2\eta')\Big|^2.\label{eq:brtv15}
\end{align}
The same expressions appear in Ref.~\cite{Ruffini:1973} [Eqs.~(11)--(12) there], and the corresponding even- and odd-parity amplitudes in Refs.~\cite{Breuer:1972cjr,Breuer:1975ncb}.

The authors did not take the limit $\delta\to0$ at fixed energy, in which $\mu^2/\delta\to9E^2$. Taking it, Eq.~\eqref{eq:brtv14} gives exactly our $\kappa_0$: the factor $\eta+\tfrac12$ is the one of Eq.~\eqref{eq:timelike} below, and the second term inside the modulus is the $m^{-1/2}$ cross term of the flux at infinity. Equation~\eqref{eq:brtv15}, instead, gives exactly $\kappa_1/4$.

We are confident in our $\kappa_1$. It rests on two independent implementations, Regge--Wheeler and Teukolsky, which agree with each other, when both are evaluated with the solutions of the Black Hole Perturbation Toolkit \cite{BHPToolkit}, to at least nine significant digits at $\ell\le5$ (Sec.~\ref{sec:setup}), and on the numerical value at $\ell=12800$, which reproduces Eq.~\eqref{eq:kodd} to about $2\times10^{-5}$ in relative terms (Sec.~\ref{sec:numerics}).

We also tested Eq.~\eqref{eq:brtv15} at finite $m$, away from the $m\to\infty$ limit, by computing exact Teukolsky fluxes at infinity for timelike orbits. The even formula is accurate to $O(m^{-1/2})$. The ratio of the exact odd flux to Eq.~\eqref{eq:brtv15} is $3.15$, $3.48$, $3.64$ and $3.77$ for $m=40$, $100$, $200$ and $400$ at $\delta=10^{-4}$. A fit $\rho_\infty+\rho_1m^{-1/2}+\rho_2m^{-1}$ to these four values gives $\rho_\infty=4.06$, and a two-term fit gives $4.05$: the ratio tends to $4$ with $O(m^{-1/2})$ corrections.

We have not traced the missing factor step by step through the 1973 derivation. Since the even formula is correct, a global error in the flux normalization is excluded, and the factor must sit in the odd-parity source or in its angular normalization. The correction matters for the total flux. With the 1973 odd term, i.e.\ with all the odd $\kappa_j$ divided by four, one would get $\kappa=0.0580$ instead of our $0.0637$, and only the latter is compatible with the numerical fit $\kappa=0.064\pm0.001$ of Ref.~\cite{Barausse:2021ryv}.

\section{Discussion}\label{sec:disc}

\subsection{Timelike orbits near the light ring}\label{sec:timelike}
Let us now go back to massive particles, which are the case of astrophysical interest, and for which Ref.~\cite{Barausse:2021ryv} also provides numerical fluxes. For a particle at $r_0=3M(1+\delta)$, with $\gamma^2=E^2/\mu^2=(1+3\delta)^2/[9\delta(1+\delta)]$, the same analysis gives, in place of Eq.~\eqref{eq:jumps}, $[\partial_r\Psi]/E=-8\pi\bar Y(1+\tfrac32\ell\delta)/(\ell M)$ and $\eta=\tfrac12+\tfrac32\ell\delta$ for $m=\ell$. Hence, at leading order in $\ell^{-1/2}$,
\begin{equation}
\Edot_{\ell\ell}\simeq\frac{\kappa_0E^2}{M^2\ell}\Big(\eta+\frac12\Big)^2\frac{F(\eta)}{F(1/2)},
\label{eq:timelike}
\end{equation}
with $F(\eta)$ the function defined after Eq.~\eqref{eq:kodd}. This is the leading term of Eq.~\eqref{eq:brtv14}. It agrees with the mean of our two numerical fluxes (the individual fluxes differ from it by $O(\ell^{-1/2})$ terms of opposite sign, at fixed $\eta$) to better than $1\%$ for $\gamma=20$ (with $\delta=1/(9\gamma^2)$) and $120\le\ell\le4800$, to $0.4$--$4\%$ for $\gamma=10$ and $30\le\ell\le1200$, and to better than $4\%$ for $\gamma=5$ and $25\le\ell\le125$.

In other words, the mass of the particle enters only through the shift of the barrier parameter $\eta$ away from $\tfrac12$. The shift grows linearly with $\ell$ and eventually switches on the exponential cut-off: since $F(\eta)\sim e^{-\pi\eta}/(\pi\sqrt{\eta/2})$ for large $\eta$ and $\tfrac32\ell\delta\simeq\ell/(6\gamma^2)$, the flux is cut off as $(\ell/6\gamma^2)^{3/2}\exp(-\pi\ell/6\gamma^2)$. The exponent is the one of the old WKB estimates; in the notation of Ref.~\cite{Barausse:2021ryv}, which uses the same $\gamma$ and $\delta$, it corresponds to the value $\pi/6$ of the coefficient called $c_1$ there. A pure exponential in $\ell/\gamma^2$, fitted to $\ell\Edot$ over a finite range of $\ell$, necessarily returns a smaller effective coefficient, which explains the value $0.42\pm0.02$ reported there.

Summing Eq.~\eqref{eq:timelike} and its $j\ge1$ analogues over $\ell$ and $m$ up to the cut-off at $\ell\sim\gamma^2$, each of the two fluxes is $\Edot_{\rm tot}M^2/E^2\simeq\kappa\ln\gamma^2+{\rm const}=0.127\ln\gamma$. Reassuringly, this is consistent with the slope $0.12\pm0.01$ (called $k_1$ there) fitted in Ref.~\cite{Barausse:2021ryv}.

\subsection{Physical interpretation}\label{sec:interp}
The divergence is ultraviolet and scale free: the power per multipole is $\kappa E^2/(M^2\ell)$, i.e.\ there is equal power per octave in $\ell$, because every multipole is excited in the same way, a fixed distance $(j+\tfrac12)\lambda_{\rm L}$ below its own quasinormal frequency, which for the dominant modes equals the damping rate of the fundamental. The near-resonance alone would not give this: without the cancellation of Sec.~\ref{sec:source} the flux per multipole would grow with $\ell$.

A heuristic picture, in the spirit of Refs.~\cite{Ruffini:1973,Hughes:1973ap}, is the following. A point particle on a circular orbit carries a near field, whose multipole $\ell$ varies along the orbit on the scale $r_0/\ell$, since $Y_{\ell\ell}\propto e^{i\ell\phi}$. On a stable orbit this near field remains bound to the particle. The light ring is unstable, and there the near field is continuously stripped away, at the rate $\lambda_{\rm L}$, and converted into radiation, half of which escapes to infinity and half falls into the hole. The near field of a point particle has a divergent energy content at short distances, i.e.\ at large $\ell$; for a null particle, after the cancellation of Sec.~\ref{sec:source}, the energy stripped per multipole falls only as $1/\ell$, and the sum diverges logarithmically.

This is also why a finite size of the source cuts the sum off, as proposed in Ref.~\cite{Barausse:2021ryv}: a source of size $d$ has no structure on scales below $d$, so it cannot excite the multipoles whose wavelength along the orbit is shorter than $d$, i.e.\ those with $\ell\gtrsim\ell_{\max}\sim r_0/d$. For a real photon $d$ is its wavelength, $\ell_{\max}\sim Er_0/\hbar$, and the flux is proportional to $\ln(EM/m_{\rm Pl}^2)$.

In Ref.~\cite{Barausse:2021ryv} we suggested that relativistic beaming was responsible for the divergence, and we used the beaming angle at the light ring to explain the equal split of the fluxes. The present analysis shows that beaming, in the form of the near-resonance between the source and the modes of the light ring, is only half of the story: a generic ultrarelativistic source in the same near-resonant condition would radiate a flux per multipole growing with $\ell$, and it is the cancellation specific to null motion that brings it down to $1/\ell$. The analysis also confirms, from a different angle, the regularization proposed there: the sum is cut off by any finite size of the source, and for a real photon that size is its wavelength.

\section{Conclusions}\label{sec:concl}
We have derived analytically a result that was previously known only from numerical solutions \cite{Barausse:2021ryv} and anticipated by the WKB estimates of the 1970's for massive particles, in the limit of vanishing rest mass at fixed energy: a massless particle on the light ring of a Schwarzschild or Kerr BH radiates GWs with a flux per multipole that decays as $1/\ell$, with the closed-form coefficient $g(a)\kappa$, $\kappa=0.063653$, equally split between infinity and the horizon, so that the total instantaneous flux diverges logarithmically. Along the way we have found that the odd-parity term of the formulae of the 1970's is too small by a factor of four, we have obtained the corrections that control the approach to the asymptotic regime, and we have reduced the whole spin dependence to the single factor $g(a)$, for which we have found a closed form.

Multiplied by the orbital period $2\pi b=6\sqrt3\,\pi M$ and summed over the two fluxes, Eq.~\eqref{eq:kappa} gives the perturbative energy loss per orbit of an ultrarelativistic body close to the critical impact parameter, $\Delta E\simeq4.2\,(E^2/M)\ln\ell_{\max}$ with half of it absorbed, in the regime $\Delta E\ll E$. This is an input for the discussion of how much energy ultrarelativistic black-hole collisions can radiate \cite{Berti:2010gx,Sperhake:2012me,Gundlach:2012aj,Page:2022}; there, the fraction of the radiated energy that is absorbed tends to $1/2$ only logarithmically in $\ell_{\max}$ (Sec.~\ref{sec:fluxes}).

The large-$\ell$ formulae, Eqs.~\eqref{eq:result} and \eqref{eq:g} with the corrections of Sec.~\ref{sec:fluxes}, give the asymptotic behavior that numerical mode sums near the light ring must reproduce, and can be used to extend such sums, whose convergence is slow \cite{Akcay:2012ea,Gundlach:2012aj}. More generally, the coefficients $\kappa_j$ measure how strongly a source localized on the light ring excites the eikonal quasinormal modes, and can serve as input for calculations of second-order (quadratic) quasinormal modes in the eikonal limit \cite{Fransen:2023eqj,Kehagias:2025,Perrone:2025,Fransen:2025,Bucciotti:2025}.

To understand the orders of magnitude involved for actual photon rings, consider, as in Ref.~\cite{Barausse:2021ryv}, the millimeter photons observed by the Event Horizon Telescope around M87$^*$ \cite{Akiyama:2019cqa}, some of which have orbited the light ring. Their wavelength cuts the sum at $\ell_{\max}\sim Er_0/\hbar=2\pi r_0/\lambda\sim10^{17}$ (for $\lambda=1.3$~mm and $M=6.5\times10^9M_\odot$), and Eq.~\eqref{eq:kappa} gives a fractional energy loss $\Delta E/E\simeq4.2\,(E/M)\ln\ell_{\max}\sim10^{-77}$ per orbit, with $E/M\simeq10^{-79}$. As already concluded there, the back-reaction on the photon ring is irrelevant for BH imaging.

Two further directions are the near-extremal limit, where the barrier flattens and the asymptotic regime recedes to large $\ell$, and horizonless compact objects with a stable light ring \cite{Cardoso:2014sna}, where a potential well replaces the barrier. There a photon on the ring would resonantly excite the long-lived trapped modes of the well, and the same construction would give the rate at which it feeds energy into them, an input for the nonlinear instability of such objects discussed in Refs.~\cite{Cardoso:2014sna,Keir:2014oka,Cunha:2022gde}.

\begin{acknowledgments}
I thank V.~Cardoso, E.~Berti, S.~A.~Hughes and G.~Khanna for discussions over many years on this problem. This work made use of the Black Hole Perturbation Toolkit \cite{BHPToolkit}. I acknowledge support from the European Union's Horizon ERC Synergy Grant ``Making Sense of the Unexpected in the Gravitational-Wave Sky'' (Grant No.\ GWSky--101167314). This work has been supported by the Agenzia Spaziale Italiana (ASI), Project n.~2024-36-HH.0, ``Attivit\`a per la fase B2/C della missione LISA''.

During the preparation of this work the author used Claude (Anthropic, Claude Fable 5.1, via Claude Code) to carry out and check symbolic and numerical calculations, to write and test the companion Mathematica and Python codes, to search the literature, and to draft and edit the text under the author's direction. All derivations, numerical results and statements were reviewed and verified by the author, who takes full responsibility for the content of the manuscript.
\end{acknowledgments}

\end{document}